\documentclass[12pt,preprint,psfig]{aastex}

\def\ltsima{$\; \buildrel < \over \sim \;$}
\def\lsim{\lower.5ex\hbox{\ltsima}}
\def\gtsima{$\; \buildrel > \over \sim \;$}
\def\gsim{\lower.5ex\hbox{\gtsima}}

\begin{document}
\title{On the Probability Distribution of Cosmological Microlensing Optical Depths}

\author{J. S. B. Wyithe\altaffilmark{1,2}, E. L. Turner\altaffilmark{1}}

\altaffiltext{1}{Princeton University Observatory, Peyton Hall, Princeton, NJ 08544, USA}

\altaffiltext{2}{School of Physics, The University of Melbourne, Parkville, Vic, 3052, Australia}

\begin{abstract}
\noindent It is conventional to calculate the probability of microlensing for
a cosmologically distant source based on the Press-Gunn approximation
that the lensing objects are uniformly and randomly distributed in
the intervening space with a constant comoving density.  We here
investigate more realistic cosmological microlensing statistics by
considering the strong spatial clustering of likely lensing objects with
each other in galaxies and their association with the clumps of dark
matter that make up the massive halos of galaxies.  Both cases in which
microlensing objects are distributed like the observed starlight in galaxies
and ones in which the extended massive halos themselves are also composed
of compact massive objects capable of acting as microlenses are
investigated.  The distribution of microlensing optical depth $\kappa$
along randomly chosen sight lines is calculated as is the conditional
distribution of $\kappa$ along sight lines near one which is strongly
microlensed.  Illustrative magnification biases are also considered.
These distributions allow us to calculate both the probability that a
high redshift source will be microlensed in the various scenarios and
the likely local $\kappa$ (averaged over nearby sight lines) at which
such microlensing events will occur.  Our overall result is that
the Press-Gunn approximation is a useful order-of-magnitude approximation
if the massive halos of galaxies are made of dark compact objects but
that {\it it fails badly and can be qualitatively misleading in the more likely
case in which only the ordinary stellar populations of galaxies are the
dominant source of cosmological microlensing events.}
In particular, we find that 
microlensing by stars is limited to of order 1 percent of high redshift
sources at any one time. 
Furthermore, even though only a small fraction of high redshift sources
are multiply-imaged (by galaxies), it is these sources that are most likely to be
microlensed by stars. 
Consequently, microlensing by stars is usually observed at $\kappa$'s near
1 where the simple isolated point mass lens approximation is not appropriate.
However, if CDM halos are composed of condensed objects, then more than 10
percent of high 
redshift sources are microlensed at any given time. The vast majority of
these sources 
are not multiply-imaged, and have $\kappa$'s smaller than 0.01.

\end{abstract}


\clearpage

\section{Introduction}

\noindent The mean optical depth to strong gravitational lensing by compact objects distributed uniformly through the universe was calculated by Press \& Gunn (1973) nearly three decades ago. The resulting probabilities are often quoted when discussing the chance that a high redshift source will be microlensed. In this paper we compute the distributions of microlensing optical depths obtained when compact objects are distributed as star light or in the dark halos of galaxies, and compare the distributions obtained to Press \& Gunn's uniform result.

Cosmological microlensing was first discussed by Chang \& Refsdal~(1979), Gott~(1981) and Young~(1981) who pointed out that the relative motion between stars in other galaxies and background sources should cause variability in the observed flux, providing an avenue to investigate an otherwise inaccessible mass regime. This phenomenon was subsequently observed in the quadruple gravitational lens Q2237+0305 (Irwin et al.~1989; Corrigan et al.~1991). The idea that most quasar variability results from microlensing due to cosmologically distributed compact objects has since been championed by Hawkins~(1993). While microlensing is most readily observed through flux variability, there are at least three other characteristic signatures. Firstly; microlensing can be detected through differential magnification of emission regions having different scales. For example, in a thorough study Dalcanton et al.~(1994) computed the effect of microlensing by randomly distributed compact objects on the equivalent widths of quasar lines. From a lack of variance in equivalent widths between high and low redshift samples, they concluded that stellar mass compact objects cannot close the universe. Secondly; microlensing of gamma ray bursts may be identified through observation of a burst that repeats. This phenomenon will arise where the burst duration is shorter than the characteristic time delay between the microlensed images. The available catalogs have been searched for lensed bursts, though none have identified (Marani et al.~1999). Using the null result, limits have been placed on the density of randomly distributed compact objects in various mass ranges between $10^{-16}$ and $10^{6}M_{\odot}$. Finally; mirolensing broadens the observed absolute brightness distribution of standard candles. Microlensing of high redshift type Ia supernovae has been discussed by Metcalf \& Silk~(1999) and Wang~(1999), and latter by M$\ddot{\mbox{o}}$rtsell, Goodbar \& Bergstr$\ddot{\mbox{o}}$m~(2001) and Minty, Heavens \& Hawkins~(2001) with emphasis on the utility of forthcoming survey samples.

In this paper we compute the probability that an image of a high-redshift source is subject to a microlensing optical depth $\kappa$. We also find the probability that a source is microlensed, as well as the conditional probability distribution of $\kappa$'s near the lines of sight to strongly microlensed sources. In Sec.~\ref{stars} we consider microlensing by stars, and in Sec.~\ref{secNFW} microlensing by dark compact objects in galactic halos. The results from these calculations are discussed in Sec.~\ref{results}. An important objective of the paper is to quantify the suitability of the assumption of randomly distributed objects. We show that the assumption is reasonable if CDM halos are composed of compact objects acting as microlenses, but is inapplicable for microlensing by stars.
Unless stated otherwise, we assume a filled beam cosmology having $\Omega=0.3$, $\Lambda=0.7$ and $h=0.65$ throughout the paper.

\section{The Probability of Microlensing by Stars}
\label{stars}

\subsection{stellar microlens distributions}

\noindent In this section we describe the probability of microlensing due to the stellar populations of galaxies. The calculations assume a universe populated by isolated (meaning no more than one lensing galaxy along any line of sight) elliptical and spiral galaxies having luminosities that are distributed according to Schechter functions of the form
\begin{equation}
\frac{dn}{dL} = \frac{n_*}{L_*}\left(\frac{L}{L_*}\right)^{\alpha}e^{-(\frac{L}{L_*})}.
\end{equation}
We also assume that galaxies obey the Faber-Jackson (1976)
\begin{equation}
\frac{L}{L_*}=\left(\frac{\sigma}{\sigma_*}\right)^{\gamma},
\end{equation}
and Tully-Fisher (1977)
\begin{equation}
\frac{L}{L_*}=\left(\frac{v_c}{v_{c*}}\right)^{\gamma}
\end{equation}
 relations, where $\sigma$ and $v_c$ are the central velocity dispersions of ellipticals and rotational velocities of spirals respectively.
The co-moving density of galaxies is assumed constant with values of $n_{*e}=(1.2\pm0.21)\times 10^{-2} h^3Mpc^{-3}$ for elliptical and S0 galaxies and $n_{*s}=(1.5\pm0.21)\times 10^{-2} h^3Mpc^{-3}$ for spirals (Marzke, Geller, Huchra \& Corwin 1994). The velocity dispersion of an $L_*$ elliptical galaxy is taken to be $\sigma_*=220\,km\,sec^{-1}$, as is the rotational velocity of an $L_*$ spiral ($v_{c_*}=220\,sec^{-1}$). The constants $\alpha$ and $\gamma$ are given the values $\alpha=-1$ and $\gamma=4$ for ellipticals, and $\alpha=-0.81$ and $\gamma=4$ for spirals. 

The light distributions of elliptical galaxies and of the bulges of spiral galaxies are well described by the de~Vaucouleurs profile. Assuming a constant mass-to-light ratio ($\Upsilon$), the surface mass density is 
\begin{equation}
\Sigma_{e}(r)= \Upsilon\hspace{2mm} \Sigma_0 10^{3.33\left(1-\left(\frac{r}{R_0}\right)^{\frac{1}{4}}\right)}.
\end{equation}
The profile has a characteristic radius $R_0$ (in $pc$) and density $\Sigma_0$ (in $kg\,m^{-2}$) given approximately by the empirical relations (Djorgovski \& Davis 1987)
\begin{equation}
\label{dv1}
\log{(h R_0)}= \frac{\log{(\sigma)}-1.632}{0.204} 
\end{equation}
and 
\begin{equation}
\label{dv2}
\log{(\Sigma_0 h^{-2})} = \log{(0.366)} + \frac{(15.06+5.48-\mu_{av})}{2.51}
\end{equation}
where 
\begin{equation}
\mu_{av} = 2.25\log{(R_0)} + 11.23.
\end{equation}
For computation of the profiles of bulges in spiral galaxies, we take the 1-dimensional velocity dispersion to be $\sigma=v_c/\sqrt{2}$.

The surface densities of thin disks in spiral galaxies (normalized to the Milky Way) can be described by a Kuzmin~(1956) disk with surface mass-density of the form (Keeton \& Kochanek 1998) 
\begin{equation}
\Sigma_{d}=0.077\frac{v_c^2}{r_d}\left(1+(\frac{r}{r_d})^2\right)^{-\frac{3}{2}}\,kg\,m^{-2},
\end{equation}
where $r_d$ is the scale length of the disk which we assume scales with $\sqrt{L}$ (Bartelmann 2000)
\begin{equation}
r_d = 3500\left(\frac{v_c}{v_{c*}}\right)^2.
\end{equation}

We assume that the distributions of stars in ellipticals and spiral bulges have spherical symmetry, and that stellar disks in spirals are circular. We also assume that the mass to light ratio is constant with radius and takes the same value for spiral bulges and elliptical galaxies. $\Upsilon$ is therefore related to the cosmological density of stars $\Omega_*$ by 
\begin{equation}
\Upsilon=\frac{\Omega_*\rho_c h^{-3} - \rho_{*d}}{\rho_{*b} + \rho_{*e}}.
\end{equation}
where 
\begin{equation}
\rho_* = \int_0^{\infty} d\sigma \frac{dn}{d\sigma} M(\sigma),
\end{equation}
$\rho_c$ is the critical density for a bound universe, and $M(\sigma)$ is the mass of a component with velocity dispersion (or circular velocity) $\sigma$. Recent CMB measurements (Netterfield et al. 2001, BOOMERANG experiment) imply a cosmic baryon density of $0.019<h^2\Omega_b<0.026$. Accordingly, we take 0.02 as the maximum value for $\Omega_*$.

In contrast to the observed light distribution, observations of flat rotation curves imply that the total mass distribution is isothermal. We assume the total mass distribution to be spherically symmetric for both elliptical and spiral galaxies. The surface mass density of a spherical singular isothermal halo with central velocity dispersion $\sigma_{h}=\sqrt{3/2}\,\sigma$ at radius $\xi$ is
\begin{equation}
\Sigma_{SIS} = \frac{\sigma_h^2}{2G\xi}
\end{equation}
where $G$ is Newtons constant. The corresponding Einstein radius is
\begin{equation}
\xi_0 = 4\pi\left(\frac{\sigma_h}{c}\right)^2\frac{D_d D_{ds}}{D_s}
\end{equation}
where $c$ is the speed of light, $D_s$ and $D_d$ are the angular diameter distances of the source (at red-shift $z_s$) and lens (at red-shift $z_d$) respectively, and $D_{ds}$ is the lens-source angular diameter distance. 

In summary, we assume that the bend angle due to gravitational lensing for both elliptical and spiral galaxies is that of the singular isothermal sphere. However, we assume the surface mass density of microlenses in elliptical/S0 galaxies to be described by the de~Vaucouleurs profile, while in spiral galaxies the microlenses are distributed as the sum of a de Vaucouleurs bulge and a thin Kuzmin disk. Thus the galaxies are composed of stars (microlenses) and smooth dark matter that sum to an isothermal mass distribution. Below (Sec.~\ref{ellprob}) we describe the calculation for microlensing probabilities in elliptical/S0 galaxies in some detail. Then in Sec.~\ref{spprob} we discuss the changes necessary for the calculation of probabilities for spirals.

\subsection{elliptical/S0 galaxies}
\label{ellprob}

\subsubsection{the distribution of $\kappa$ for singly-imaged sources}

\noindent The microlensing optical depth $\kappa$ is defined to be the ratio of surface mass density $\Sigma$ to the critical density 
\begin{equation}
\Sigma_{crit} = \frac{c^2}{4\pi G}\frac{D_s}{D_d D_{ds}},
\end{equation}
and the optical depth in stars at a radius $\xi$ is therefore
\begin{equation}
\kappa_{*}= \Upsilon \frac{\Sigma_0}{\Sigma_{crit}} 10^{3.33\left(1-\left(\frac{\xi}{R_0}\right)^{\frac{1}{4}}\right)}.
\end{equation}
The smooth (dark) component of dimensionless surface mass density that maintains the isothermal mass distribution can be written
\begin{equation}
\kappa_c = \frac{\Sigma_{SIS}}{\Sigma_{crit}}-\kappa_{*}.
\end{equation}
The presence of continuous matter modifies the microlensing probability due to $\kappa_*$, and the effective optical depth to microlensing due to $\kappa_*$ is 
\begin{equation}
\kappa(\xi)=\frac{\kappa_*(\xi)}{|1-\kappa_c(\xi)|}
\end{equation}
which is not monotonic in $\xi$. $\{\xi_{\kappa ,in},\xi_{\kappa ,out}\}$ is the set of $N_{pairs}$ pairs of radii having the property that the optical depth is larger than $\kappa$ at all $\xi_{\kappa ,in}<\xi<\xi_{\kappa ,out}$. Extending the calculation of Koopmans \& Wambsganss (2001) and following Turner, Ostriker \& Gott (1984), we find the probability that a beam from a singly-imaged (where singly-imaged refers to macrolensing due to the galaxy) source will pass through a microlensing optical depth (as seen by the observer) {\em larger} than $\kappa$ due to an elliptical galaxy having a central velocity dispersion between $\sigma$ and $\sigma+\Delta\sigma$ and a redshift between $z_d$ and $z_d+\Delta z_d$. All probabilities are calculated for sources at a single redshift $z_s$. The resulting differential cross-section is found as a function of central velocity dispersion $\sigma$, and may be written
\begin{equation}
\label{dtaudz}
\frac{d\tau}{dz_d}(\sigma,\kappa) = \frac{dn}{d\sigma}\Delta \sigma \frac{c\pi}{H_0} \frac{dH_0t}{dz}|_{z=z_d}(1+z_d)^3\sum_{i=1}^{N_{pairs}}\left(\mbox{max}(\zeta^i_{out},\xi_0)^2 - \mbox{max}(\zeta^i_{in},\xi_0)^2\right),
\end{equation}
where $\zeta$ is the unlensed impact parameter obtained from the lens equation for a singular isothermal sphere
\begin{equation}
\zeta = \xi_{\kappa}-\xi_0,
\end{equation}
and $\frac{dH_0t}{dz}$ is obtained from the expression for lookback time from the present (Carrol, Press \& Turner 1992)
\begin{equation}
\frac{dH_0t}{dz} = (1+z)^{-1}\left[(1+z)^2(1+\Omega z) - z_d(2+z)\Lambda\right]^{-\frac{1}{2}}.
\end{equation}
Since $\frac{d\tau}{dz_d}(\sigma,\kappa)$ is monotonic in $\kappa$ for all $z_d$, the probability that a beam from a singly-imaged source will pass through a microlensing optical depth between $\kappa$ and $\kappa+\Delta \kappa$ due to a galaxy between redshift $z_d$ and $z_d+\Delta z_d$ is found from the derivative of Eqn.~\ref{dtaudz} for each $\sigma$:
\begin{equation}
\label{d2taudzdk}
\frac{d^2\tau}{d z_d d\kappa}(\sigma) = \frac{d}{d\kappa}\left(\frac{d\tau}{dz_d}(\sigma,\kappa)\right).
\end{equation}

The magnification of a lensed image has the potential to boost the observed flux of a source which is intrinsically faint above the detection magnitude limit $m_{lim}$ (Gott \& Gunn~1974; Turner 1980), resulting in a bias for highly magnified images. The magnification bias for an image with co-ordinate $\xi_{\kappa}$ is 
\begin{equation}
\label{biaseqn}
B(\xi_{\kappa})\equiv\frac{N\left(<m_{lim}+\frac{5}{2}\log\left(|\mu\left(\xi_{\kappa}\right)|\right)\right)}{N(<m_{lim})}.
\end{equation}
Since $\xi(\kappa)$ is multi-valued, the magnification bias with respect to an image being subject to microlensing optical depth between $\kappa$ and $\kappa+\Delta \kappa$ due to a galaxy between redshift $z_d$ and $z_d+\Delta z_d$ with a central velocity dispersion between $\sigma$ and $\sigma+\Delta\sigma$ is
\begin{equation}
\label{bias_eq}
B_{\kappa}(z_d,\sigma) = \frac{\Sigma_{i=1}^{N_{pairs}}\left((\zeta_{in}^i)^2 B(\xi_{\kappa,in}) +  (\zeta_{out}^i)^2 B(\xi_{\kappa,out})\right)}{\Sigma_{i=1}^{N_{pairs}}\left((\zeta_{in}^i)^2 + (\zeta_{out}^i)^2\right)}
\end{equation}
Note that this neglects magnification bias due to microlensing. Including magnification bias we find the probability that a singly-imaged source will be observed through a surface mass density between $\kappa$ and $\kappa+\Delta \kappa$ due to a galaxy between redshift $z_d$ and $z_d+\Delta z_d$ for each $\sigma$:
\begin{equation}
\label{d2Pdzdk}
\frac{d^2P_{\kappa}}{d z_d d\kappa}(\sigma) = B_{\kappa}(z_d,\sigma)\frac{d^2\tau}{d z_d d\kappa}.
\end{equation}

Using Eqn.~\ref{d2Pdzdk} and the Poisson probability for a lensing event at optical depth $\kappa_{ML}$, we approximate the conditional differential probability that a singly-imaged, \emph{strongly microlensed} source will have an optical depth in stars near the line of sight of between $\kappa_{ML}$ and $\kappa_{ML}+\Delta \kappa_{ML}$ in a galaxy between redshift $z_d$ and $z_d+\Delta z_d$ for each $\sigma$:
\begin{eqnarray}
\label{d2PMLdzdk}
\nonumber
\frac{d^2P_{ML}}{d z_d d\kappa_{ML}}(\sigma) &=& (1-e^{-\kappa_{ML}})\frac{d^2P_{\kappa}}{d z_d d\kappa}|_{\kappa=\kappa_{ML}}\\
&\sim& \kappa_{ML}\frac{d^2P_{\kappa_{ML}}}{d z_d d\kappa}|_{\kappa=\kappa_{ML}}\hspace{3mm}\mbox{for}\hspace{3mm}\kappa_{ML}\ll1.
\end{eqnarray}
This ignores the contribution of shear. However, at low surface mass densities microlensing in the presence of shear is due to an ensemble of Chang-Refsdal lenses, which have similar cross-sections to point mass lenses. At normalized surface mass densities near 1, the Chang-Refsdal diamond caustic is replaced by a continuous caustic network, which has a cross-section near 1. We therefore feel this approximation to be sufficient for the current purpose.

\begin{figure*}[hptb]
\epsscale{1.}
\plotone{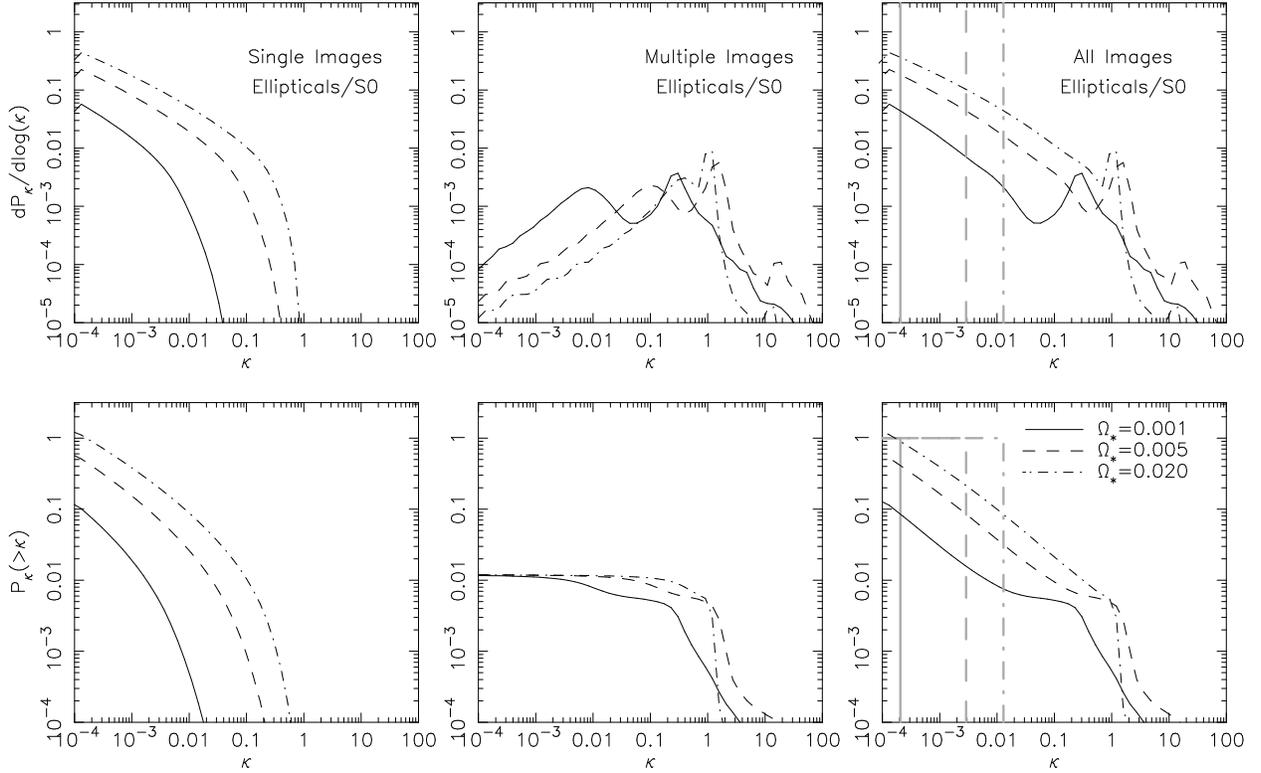}
\caption{The probability of the microlensing optical depth due to stars in elliptical/S0 galaxies: Differential (top) and cumulative (bottom) probabilities for $\kappa$ are shown. Left: Singly-imaged sources, Center: Multiply-imaged sources, Right: All images. The source redshift was $z_s=3$ and no magnification bias was assumed. Functions are shown for three values of $\Omega_*$: 0.001 (solid lines), 0.005 (dashed lines), and 0.020 (dot-dashed lines). For comparison, the light lines show the corresponding probabilities where $\Omega_{e*}$ is uniformly distributed.}
\label{SISell1}
\end{figure*}

\begin{figure*}[hptb]
\epsscale{1.}
\plotone{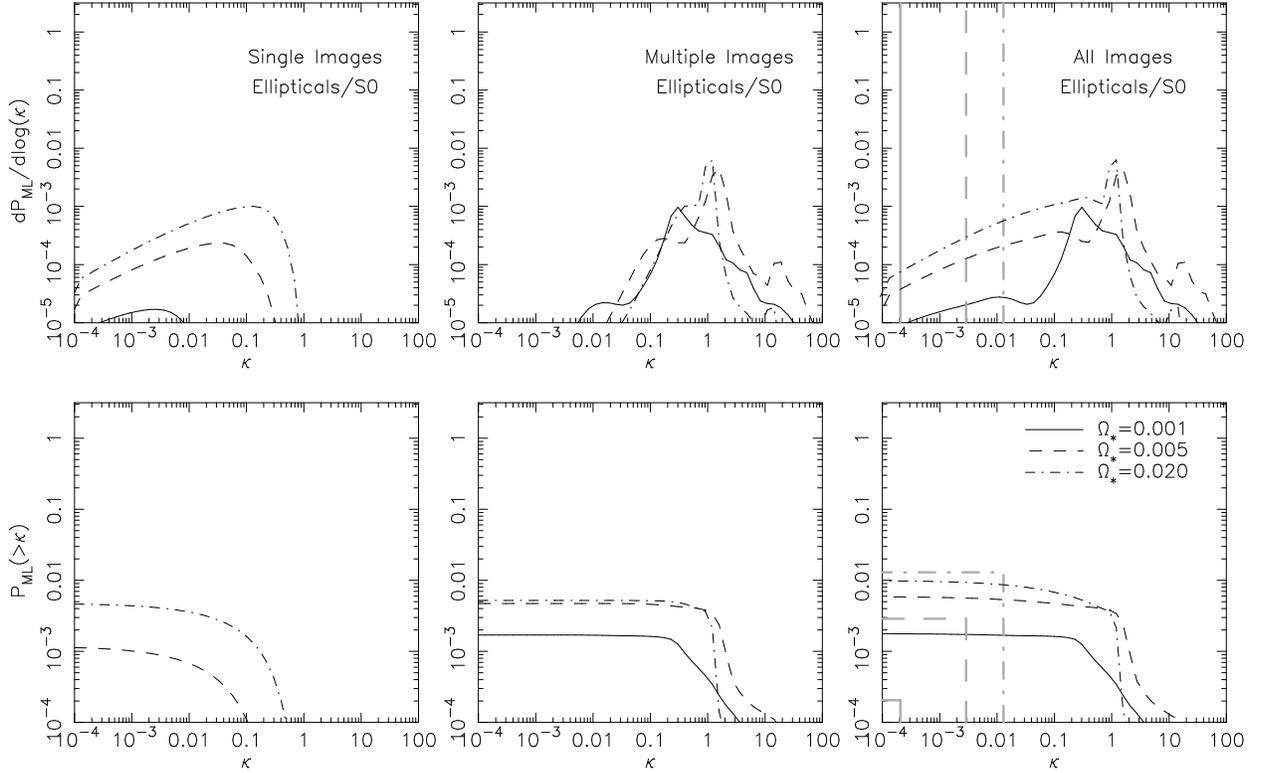}
\caption{The conditional probability of the microlensing optical depth along lines of sight near microlensed images in elliptical/S0 galaxies. Differential (top) and cumulative (bottom) probabilities for $\kappa$ are shown. Left: Singly-imaged sources, Center: Multiply-imaged sources, Right: All images. The source redshift was $z_s=3$ and no magnification bias was assumed. Functions are shown for three values of $\Omega_{*}$: 0.001 (solid lines), 0.005 (dashed lines) and 0.020 (dotted lines). For comparison, the light lines show the corresponding probabilities where $\Omega_{e*}$ is uniformly distributed.}
\label{SISell2} 
\end{figure*}

Finally, the dependences on $\sigma$ and $z_d$ are integrated out of Eqns.~\ref{d2Pdzdk} and \ref{d2PMLdzdk} yielding
\begin{equation}
\label{dPdkap}
\frac{dP_{\kappa}}{d\kappa} = \int_0^{z_s}dz_d\int_0^\infty d\sigma \frac{dn}{d\sigma}\frac{d^2P_{\kappa}}{d z_d d\kappa}(\sigma)
\end{equation}
and 
\begin{equation}
\label{dPMLdkap}
\frac{dP_{ML}}{d\kappa_{ML}} = \int_0^{z_s}dz_d\int_0^\infty d\sigma  \frac{dn}{d\sigma}\frac{d^2P_{ML}}{d z_d d\kappa_{ML}}(\sigma).
\end{equation}
To compute cumulative probabilities, differential probabilities are integrated from $\kappa$ to $\infty$ because $\frac{dP_{\kappa}}{d\kappa}\rightarrow 0$ as $\kappa\rightarrow \infty$. 
The probability that a singly-imaged source is subject to a microlensing optical depth greater than $\kappa$ is then
\begin{equation}
\label{P}
P_{\kappa}(>\kappa) = \int_{\kappa}^{\infty} d\kappa' \frac{dP_{\kappa}}{d\kappa'}.
\end{equation}
Similarly, the probability that a singly-imaged microlensed source will have an optical depth in stars near the line of sight greater than $\kappa_{ML}$ is
\begin{equation}
\label{PML}
P_{ML}(>\kappa_{ML}) = \int_{\kappa_{ML}}^{\infty} d\kappa_{ML}' \frac{dP_{ML}}{d\kappa_{ML}'}.
\end{equation}
$\frac{dP_{\kappa}}{d\kappa}$ and $P_{\kappa}(>\kappa)$, as well as $\frac{dP_{ML}}{d\kappa_{ML}}$ and $P_{ML}(>\kappa_{ML})$ due to elliptical/S0 galaxies for singly-imaged sources are plotted in the left hand panels of Figs.~\ref{SISell1} and \ref{SISell2} for $z_s=3$ and $\Omega_*=0.001$, 0.005, and 0.020. No magnification bias (i.e. $B_{\kappa}=1$) was assumed for these plots, but is discussed in Sec.~\ref{magbiassec}.

\subsubsection{the distribution of $\kappa$ for multiply-imaged sources}

\noindent The equivalent calculation for multiply-imaged (where multiply-imaged refers to macrolensing due to the galaxy) sources was made by replacing the cross-section in Eqn.~\ref{dtaudz} with
\begin{equation}
\label{dtaudz2}
\frac{d\tau}{dz_d}(\sigma,\kappa) = 2\frac{dn}{d\sigma}\Delta \sigma  \frac{c\pi}{H_0} \frac{dH_0t}{dz}|_{z=z_d}(1+z_d)^3
\sum_{i=1}^{N_{pairs}} \left\{ \begin{array}{ll}
\mbox{min}(\zeta^i_{out},\xi_0)^2 - \mbox{min}(\zeta^i_{in},\xi_0)^2    & \hspace{-1mm} \mbox{if}\hspace{1mm}\zeta_{in}^i>0\\
\mbox{min}(\zeta^i_{out},\xi_0)^2 + \left(\zeta_{in}^i\right)^2  & \hspace{-1mm} \mbox{if}\hspace{1mm}\zeta_{in}^i<0\\
\left(\zeta_{in}^i\right)^2 - \left(\zeta_{out}^i\right)^2 &\hspace{-1mm} \mbox{if}\hspace{1mm}\zeta_{out}^i<0
\end{array}\right.
\end{equation}
which is the probability that a beam from a multiply-imaged source will pass through a microlensing optical depth {\em smaller} than $\kappa$ due to a galaxy between redshift $z_d$ and $z_d+\Delta z_d$ with a central velocity dispersion between $\sigma$ and $\sigma+\Delta\sigma$. $\frac{dP_{\kappa}}{d\kappa}$ and $P_{\kappa}(>\kappa)$, as well as $\frac{dP_{ML}}{d\kappa_{ML}}$ and $P_{ML}(>\kappa_{ML})$ due to elliptical/S0 galaxies for multiply-imaged sources are plotted in the central panels of Figs.~\ref{SISell1} and \ref{SISell2} for $z_s=3$ and $\Omega_*=0.001$, 0.005, and 0.020.  Here no magnification bias was assumed, this is discussed in Sec.~\ref{magbiassec}. 

\subsubsection{the distribution of $\kappa$ for all sources}

\noindent $\frac{dP_{\kappa}}{d\kappa}$, $\frac{dP_{ML}}{d\kappa_{ML}}$, $P_{\kappa}(>\kappa)$ and $P_{ML}(>\kappa_{ML})$ due to elliptical/S0 galaxies for all sources were found by summing the pairs of differential and cumulative distributions described in the previous two subsections for singly and multiply-imaged sources. These are plotted in the right hand panels of Figs.~\ref{SISell1} and \ref{SISell2} for $z_s=3$ and $\Omega_*=0.001$, 0.005, and 0.020.

\subsection{spiral galaxies}
\label{spprob}

\noindent For spiral galaxies, the optical depth in stars at a radius $\xi$ is calculated from the sum of the bulge and disc surface mass densities 
\begin{equation}
\kappa_{*}= \Upsilon \frac{\Sigma_0}{\Sigma_{crit}} 10^{3.33\left(1-\left(\frac{\xi}{R_0}\right)^{\frac{1}{4}}\right)}+ \frac{1}{sin{(i)}}\frac{0.077}{\Sigma_{crit}}\frac{v_c^2}{r_d}\left(1+(\frac{\xi_s}{r_d})^2\right)^{-\frac{3}{2}}
\end{equation}
where 
\begin{equation}
\xi_s = \sqrt{\left(\xi \cos{(\theta)}\right)^2 + \left(\frac{1}{sin{(i)}}\xi \sin{(\theta)}\right)^2},
\end{equation}
$i$ is the inclination of the disk ($i=\frac{\pi}{2}$ is face on), and $\theta$ is the angle between the image position and the semi-major axis (subtended at the galactic center). $\Sigma_0$ and $R_0$ were calculated from Eqns.~\ref{dv1} and \ref{dv2}, with $\sigma=\frac{v_{c}}{\sqrt{3}}$.

\begin{figure*}[hptb]
\epsscale{1.}
\plotone{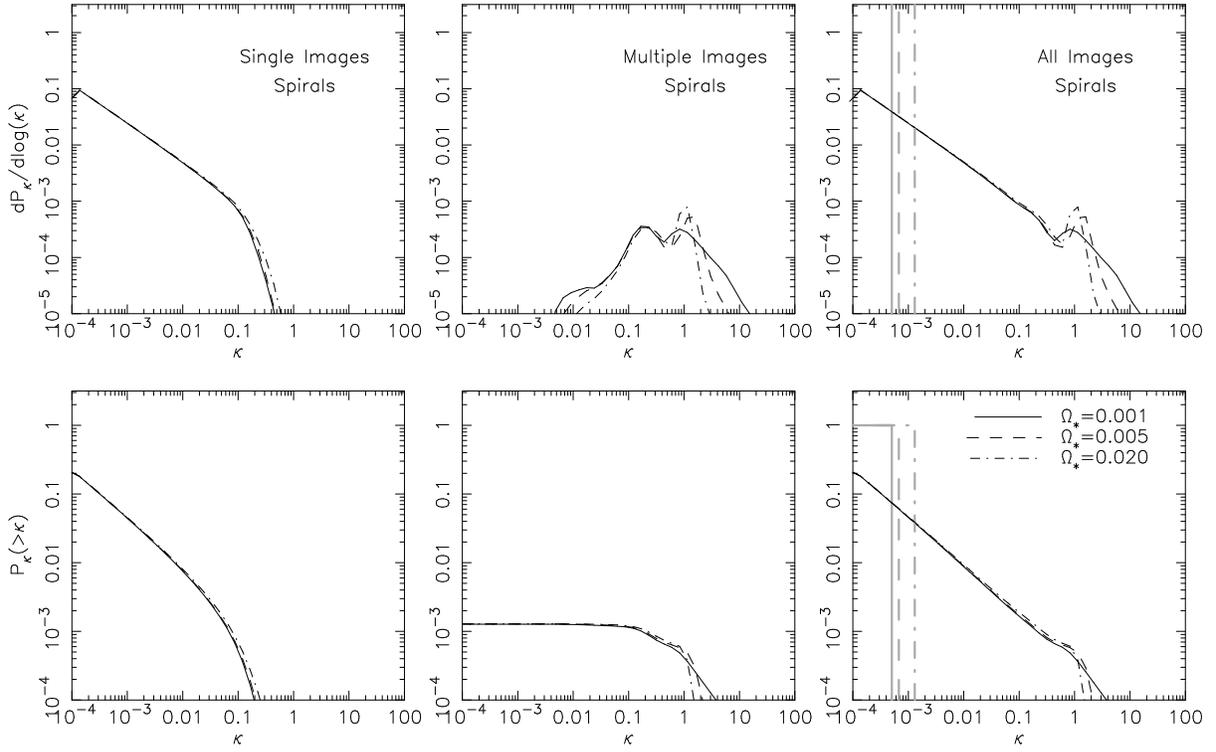}
\caption{The probability of the microlensing optical depth due to stars in spiral galaxies: Differential (top) and cumulative (bottom) probabilities for $\kappa$ are shown. Left: Singly-imaged sources, Center: Multiply-imaged sources, Right: All images. The source redshift was $z_s=3$ and no magnification bias was assumed. Functions are shown for three values of $\Omega_*$: 0.001 (solid lines), 0.005 (dashed lines) and 0.020 (dotted lines). For comparison, the light lines show the corresponding probabilities where $\Omega_{s*}$ is uniformly distributed.}
\label{SISsp1}
\end{figure*}

\begin{figure*}[hptb]
\epsscale{1.}
\plotone{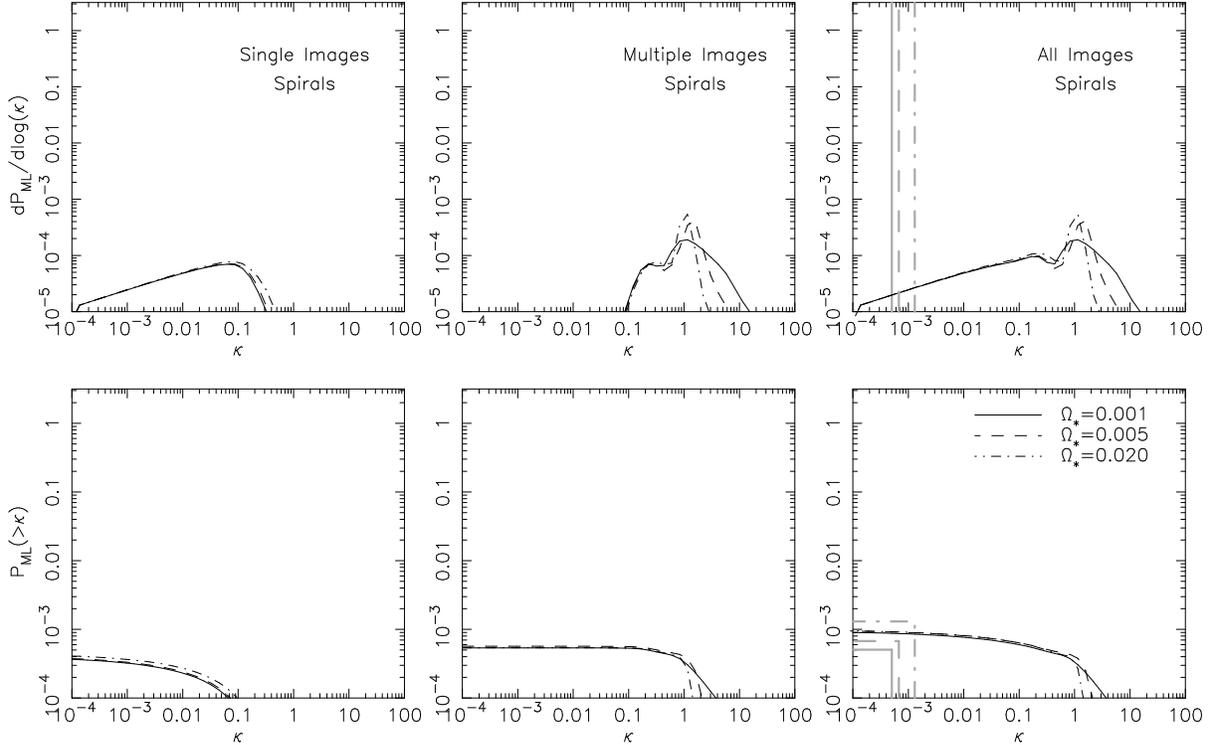}
\caption{The conditional probability of the microlensing optical depth along lines of sight near microlensed images in spiral galaxies. Differential (top) and cumulative (bottom) probabilities for $\kappa$ are shown. Left: Singly-imaged sources, Center: Multiply-imaged sources, Right: All images. The source redshift was $z_s=3$ and no magnification bias was assumed. Functions are shown for three values of $\Omega_*$: 0.001 (solid lines), 0.005 (dashed lines) and 0.020 (dot-dashed lines). For comparison, the light lines show the corresponding probabilities where $\Omega_{s*}$ is uniformly distributed.}
\label{SISsp2} 
\end{figure*}

The bias $B_{\kappa}(\sigma,\kappa,i,\theta)$ (Eqn.~\ref{bias_eq}) and cross-section $\frac{d\tau}{dz_d}(\sigma,\kappa,i,\theta)$ (Eqn.~\ref{dtaudz}) are computed as a function of $i$ and $\theta$. Since we have assumed that the bend angle is that of an isothermal sphere, the inclination and orientation averaged bias and cross-section are 
\begin{equation}
B_{\kappa}(z_d,\sigma) = \frac{\pi^2}{4}\int_0^{\frac{\pi}{2}}di \int_0^{\frac{\pi}{2}}d\theta B_{\kappa}(z_d,\sigma,i,\theta)
\end{equation} 
and 
\begin{equation}
\frac{d\tau}{dz_d}(\sigma,\kappa) = \frac{\pi^2}{4}\int_0^{\frac{\pi}{2}}di \int_0^{\frac{\pi}{2}}d\theta \frac{d\tau}{dz_d}(\sigma,\kappa,i,\theta).
\end{equation} 
Figs.~\ref{SISsp1} and \ref{SISsp2} show $\frac{dP_{\kappa}}{d\kappa}$ and $P_{\kappa}(>\kappa)$, as well as $\frac{dP_{ML}}{d\kappa_{ML}}$ and $P_{ML}(>\kappa_{ML})$ for singly-imaged (left hand panels), multiply imaged (center panels) and all (right-hand panels) sources due to spiral galaxies. The source redshift was $z_s=3$ and $\Omega_*=0.001$, 0.005 and 0.020. No magnification bias was assumed for these plots.

\subsection{variation with source redshift}
\label{sourcez}

\begin{figure*}
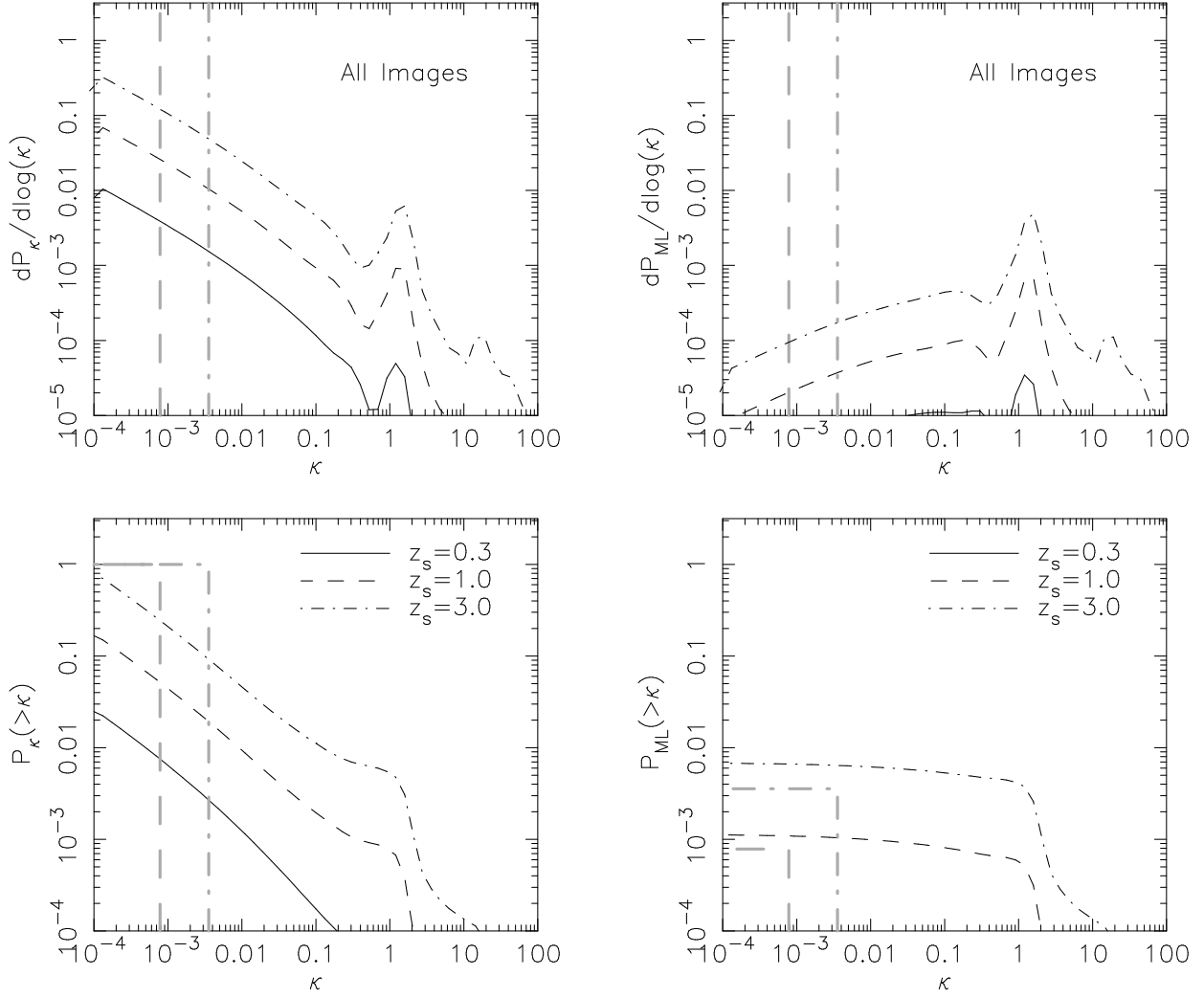

\epsscale{1.}
\plotone{f5.epsi}
\includegraphics{SIS_Zs2.ps}
\caption{The dependence of the probability of microlensing by stars on source redshift: Left: Differential (top) and cumulative (bottom) probabilities for the microlensing optical depth $\kappa$. Right: Conditional differential (top) and cumulative (bottom) probabilities for the $\kappa$ along lines of sight near microlensed images. $\Omega_*$ was 0.005, $\Lambda$ was 0.7 ($\Omega + \Lambda=1$), and no magnification bias was assumed. Functions are shown for three values of $z_s$: 0.3 (solid lines), 1.0 (dashed lines) and 3.0 (dot-dashed lines). For comparison, the light lines show the corresponding probabilities where $\Omega_*$ is uniformly distributed.}
\label{SIS_Zs} 
\end{figure*}

\noindent Turner, Ostriker \& Gott~(1984) showed that the probability of multiple imaging by isothermal galaxies is a sensitive function of source redshift ($\propto z_s^3$ for $z_s\ll1$). We therefore expect that microlensing is more likely for high redshift sources.  Fig.~\ref{SIS_Zs} shows $\frac{dP_{\kappa}}{d\kappa}$, $\frac{dP_{ML}}{d\kappa_{ML}}$, $P_{\kappa}(>\kappa)$ and $P_{ML}(>\kappa_{ML})$ for microlensing by the combined elliptical and spiral galaxy populations for all sources at redshifts $z_s=$ 0.3, 1.0 and 3.0. No magnification bias was assumed, and $\Omega_*=0.005$.

\subsection{variation with cosmology}
\label{sourcez}

\begin{figure*}
\epsscale{1.}
\plotone{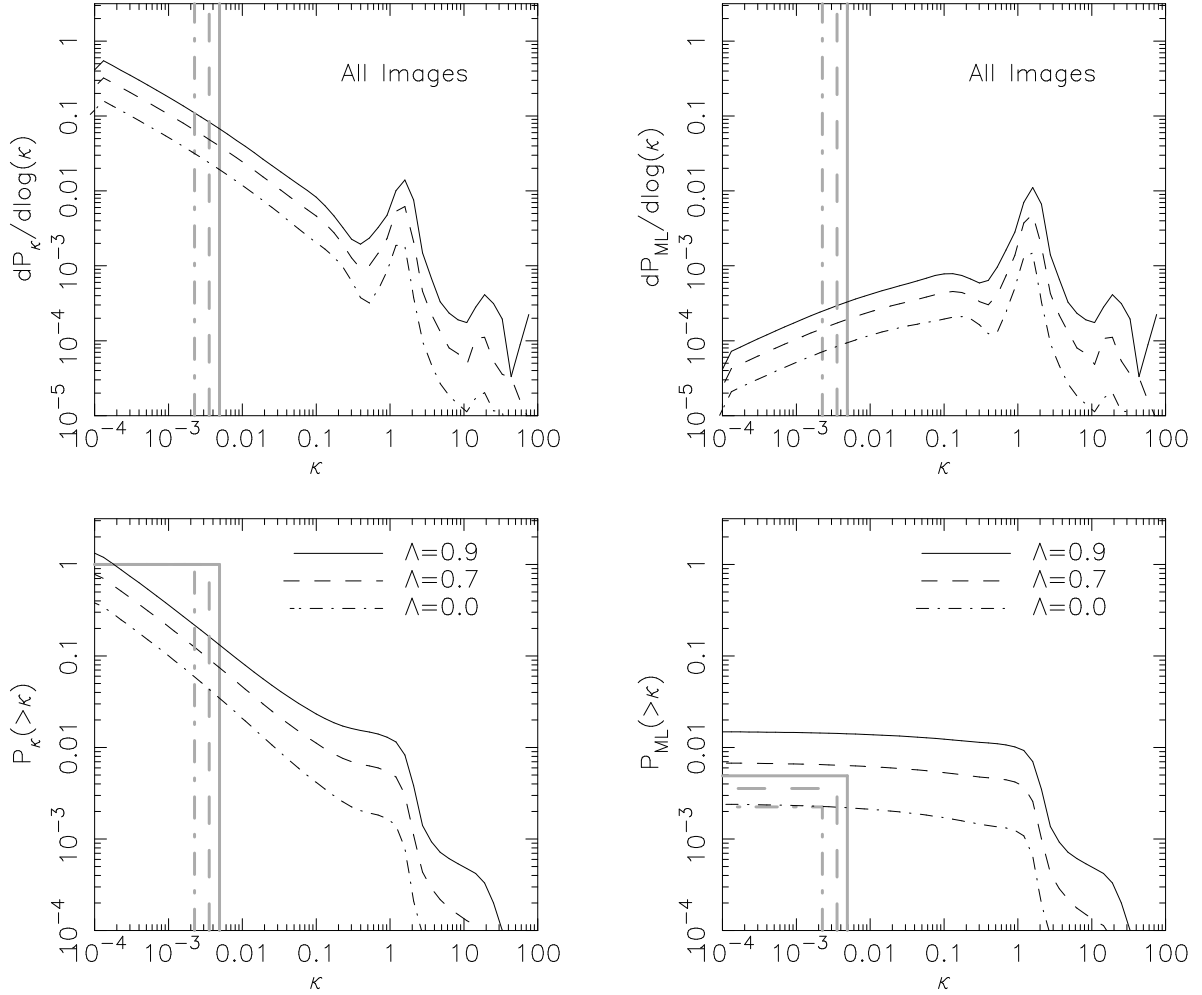}
\caption{The dependence of the probability of microlensing by stars on cosmology: Left: Differential (top) and cumulative (bottom) probabilities for the microlensing optical depth $\kappa$. Right:  Conditional differential (top) and cumulative (bottom) probabilities for the $\kappa$ along lines of sight near microlensed images. $\Omega_*$ was 0.005, the source redshift was $z_s=3.0$, and no magnification bias was assumed. All cosmologies had $\Omega+\Lambda=1$. Functions are shown for three values of $\Lambda$: 0.9 (solid lines), 0.7 (dashed lines), 0.5 (dot-dashed lines) and 0.0 (dotted lines). For comparison, the light lines show the corresponding probabilities where $\Omega_*$ is uniformly distributed.}
\label{SIS_cos} 
\end{figure*}

\noindent Turner~(1990) and Fukugita, Futamase \& Kasai (1990) showed that in a flat universe, the presence of a non-zero cosmological constant significantly increases the frequency of multiply imaged gravitationally lensed quasars. We therefore expect a corresponding increase in the probability of microlensing. Fig.~\ref{SIS_cos} shows $\frac{dP_{\kappa}}{d\kappa}$, $\frac{dP_{ML}}{d\kappa_{ML}}$, $P_{\kappa}(>\kappa)$ and $P_{ML}(>\kappa_{ML})$ for microlensing by the combined elliptical and spiral galaxy populations for all sources assuming $\Omega+\Lambda=1$ and $\Lambda=0.9$, 0.7 and 0.0. No magnification bias was assumed, the source red-shift was $z_s=3.0$, and $\Omega_*=0.005$. 

\subsection{variation with magnification bias}
\label{magbiassec}

\noindent To compute the magnification bias, we use the number magnitude relation described by Kochanek (1996)
\begin{equation}
\frac{dN}{dm}=N_010^{\alpha_m m},
\label{numbercount}
\end{equation}
where $N_0$ is a normalizing constant and $\alpha_m$ is the logarithmic slope. To simplify our calculations and to keep the conclusions as general as possible, we assume that this form is valid for all sources at redshift $z_s$ fainter than the survey depth. As an example, for quasars, and based on data from Boyle, Shanks \& Peterson (1988) and Hartwick \& Schade (1990), Kockanek (1996) finds that at $z_s=3$, $\alpha_m\sim0.27$ below $m_B\sim19$. 

\begin{figure*}
\epsscale{1.}
\plotone{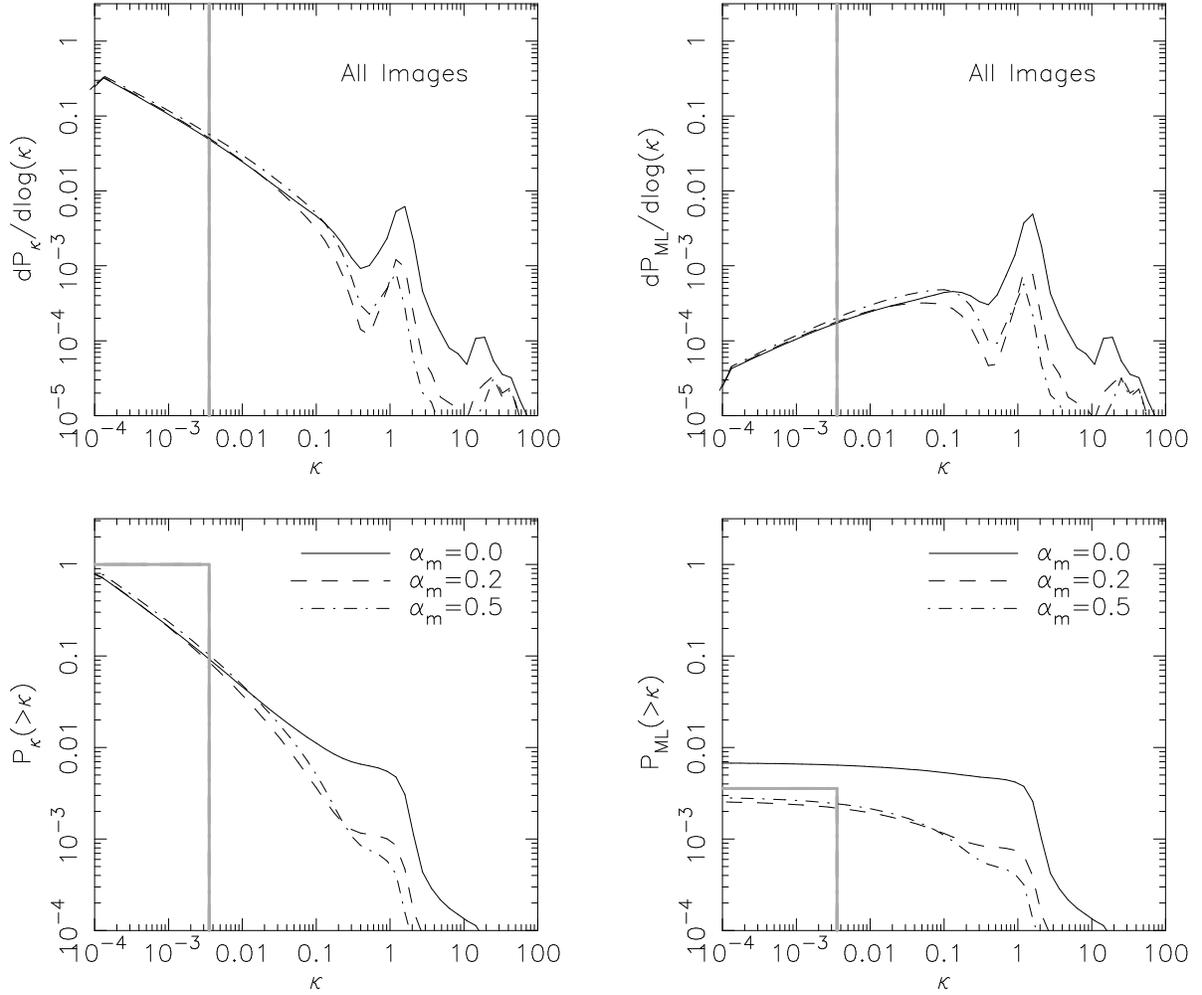}
\caption{The dependence of the probability of microlensing by stars on the magnification bias. Left: Differential (top) and cumulative (bottom) probabilities for the microlensing optical depth $\kappa$. Right:  Conditional differential (top) and cumulative (bottom) probabilities for the $\kappa$ along lines of sight near microlensed images. $\Omega_*$ was 0.005, and a source redshift of $z_s=3.0$ was assumed. Functions are shown for three values of $\alpha_m$: 0.0 (unbiased, solid lines), 0.2 (dashed lines) and 0.5 (dot-dashed lines). For comparison, the light lines show the corresponding probabilities where $\Omega_*$ is uniformly distributed.}
\label{SIS_BIAS} 
\end{figure*}

In the present case of an isothermal mass distribution Eqn.~\ref{biaseqn} reduces to
\begin{equation}
B=\left(\left|\frac{\xi_{\kappa}}{|\xi_{\kappa}|-\xi_0}\right|\right)^{\frac{5}{2}\alpha_m}.
\end{equation}
Fig.~\ref{SIS_BIAS} shows $\frac{dP_{\kappa}}{d\kappa}$, $\frac{dP_{ML}}{d\kappa_{ML}}$, $P_{\kappa}(>\kappa)$ and $P_{ML}(>\kappa_{ML})$ for microlensing by the combined elliptical and spiral galaxy populations for all sources at redshift $z_s=$3.0, and assuming $\Omega_*=0.005$. Magnification biases were calculated assuming 3 different values of $\alpha_m$: 0.0 (unbiased, solid lines), 0.2 (dashed lines) and 0.5 (dot-dashed lines).

\section{The Probability of Microlensing by CDM Halos}
\label{secNFW}

\noindent In this section we assume that dark matter is in the form of compact objects and calculate the probability of microlensing in the dark halos around galaxies. We assume that dark matter halos are described by the (NFW) profile of Navarro, Frenk \& White (1995, 1996, 1997). Thus both the bend angle due to gravitational lensing, and the surface mass density of microlenses are described by the projection of the NFW profile. The halos are assumed to be isolated (meaning no more than one lensing galaxy along any line of sight), and when combined, to account for a cosmological density $\Omega_o\rho_{crit}$. 

Following Bullock et al. (2001), the NFW profile has the following two parameter functional form for space density as a function of radius $r$ 
\begin{equation}
\label{NFWprofile}
\rho_{NFW}(r) = \frac{\rho_s}{\frac{r}{r_s}(1+\frac{r}{r_s})^2},
\end{equation}
which yields a microlensing optical depth in compact objects of (Bartelmann 1996)
\begin{equation}
\kappa_{co} = \frac{1}{\Sigma_{crit}}\frac{2\rho_s r_s}{\left(\frac{\xi}{r_s}\right)^2-1}f\left(\frac{\xi}{r_s}\right),
\end{equation}
where 
\begin{equation}
f(x)=\left\{ \begin{array}{ll}
1 - \frac{2}{\sqrt{x^2-1}}\tan^{-1}\sqrt{\frac{x-1}{x+1}}  & \mbox{if}\:\:\:x>1\\
1 - \frac{2}{\sqrt{1-x^2}}\tanh^{-1}\sqrt{\frac{1-x}{1+x}}  & \mbox{if}\:\:\:x<1\\
0                                                          & \mbox{if}\:\:\:x=1
\end{array}\right.
\end{equation}
and $\rho_s$ and $r_s$ are the characteristic density and radius.  The concentration of the halo is defined as
\begin{equation}
C_{vir} = \frac{r_{vir}}{r_s},
\end{equation}
where $r_{vir}$ is the radius containing an over-density of $\Delta_{vir}$. In flat cosmologies, $\Delta_{vir}$ is dependent on $\Omega$ and at redshift zero is approximated by (Bryan \& Norman 1997)
\begin{equation}
\Delta_{vir}\sim\frac{18\pi^2+82(\Omega-1)-39(\Omega-1)^2}{\Omega}.
\end{equation}
The virial mass of the halo $M_{vir}$ is 
\begin{equation}
M_{vir}=\frac{4\pi}{3}\Delta_{vir}\rho_c r_{vir}^3,
\end{equation}
and to complete the relations between parameters, the characteristic density is
\begin{equation}
\label{rhos}
\rho_s=\frac{M_{vir}}{4\pi r_s^3}\left(\log{(1+C_{vir})}-\frac{C_{vir}}{1+C_{vir}}\right)^{-1}.
\end{equation}
Bullock et al.~(2001) studied the density profiles of a large sample of halos in a high resolution N-body simulation. They assumed a cosmology having $\Omega=0.3$ and $\Lambda=0.7$, and found the following mean relationships:
\begin{equation}
\log{\left(\frac{M_{vir}}{h^{-1}}\right)} \sim 3.4\log{\left(\frac{V_{max}}{km\,sec^{-1}}\right)} + 4.3
\end{equation}
where $V_{max}$ is the maximum orbital velocity (occurring at $r\sim2.16r_s$) and 
\begin{equation}
\label{C_vir}
C_{vir}\sim9\left(\frac{M_{vir}}{2\times10^{13}h^{-1}M_{\odot}}\right)^{-0.13}.
\end{equation}
Taking $V_{max}=\sqrt{2}\sigma$, we find the density profile as a function of $\sigma$ using Eqns.~\ref{NFWprofile}-\ref{C_vir}. Note that the assumption of an un-evolving lens population is supported in this case by Bullock et al. who find that $r_s$ is quite insensitive to redshift. The NFW halo does not contain finite mass. We find that cylinders of radius $R_{max}\sim1 R_{vir}$ around all halos contain enough mass to account for the density $\Omega\rho_{crit}$, and therefore assume $\kappa_{co}$ to equal zero beyond $R_{max}$. This assumption only affects probabilities for $\kappa<10^{-3}$.

The lens equation for an NFW profile (Bartelmann 1996; Li \& Ostriker 2001) is 
\begin{equation}
\label{NFWeqn}
\eta = \xi - \alpha_s \frac{g\left(\frac{\xi}{r_s}\right)}{\xi}
\end{equation}
where $\eta$ is the source position (defined in the source plane),
\begin{equation}
g(x)=1-f(x)+\ln{\left(\frac{x}{2}\right)}
\end{equation}
and 
\begin{equation}
\alpha_s = \frac{4\rho_s r_s}{\Sigma_{crit}}.
\end{equation}    
The critical image radius ($\xi_{crit}$) for multiple imaging is the solution of 
\begin{equation}
\frac{d\eta}{d\xi}-\frac{D_d}{D_s}=0
\end{equation}
which yields from Eqn.~\ref{NFWeqn} the critical position ($\eta_{crit}$) inside which a source is multiply-imaged, and the resulting critical impact parameter 
\begin{equation}
\xi_{mult} = \frac{D_d}{D_s}\eta_{crit}.
\end{equation}

We have repeated the calculations of the previous section for microlensing due to compact objects in dark matter halos. Only the cross-sections are described below, as the procedure is analogous to that for microlensing by stars, with the following differences: Firstly, $\kappa$ is more simply defined by
\begin{equation}
\kappa = \kappa_{co},
\end{equation}
and decreases monotonically with $\xi$. Because $\xi(\kappa)$ is single valued, $B_{\kappa}=B(\xi_{\kappa})$. Secondly, the un-lensed impact parameter is now given by 
\begin{equation}
\zeta = \frac{D_d}{D_s}\eta(\xi_\kappa)
\end{equation}
together with Eqn.~\ref{NFWeqn}.

\subsection{the distribution of $\kappa$ for singly-imaged sources}

\noindent The probability that a beam from a singly-imaged source will pass through a microlensing optical depth larger than $\kappa$ due to a galaxy halo with a velocity dispersion between $\sigma$ and $\sigma+\Delta\sigma$ and a redshift between $z_d$ and $z_d+\Delta z_d$ is 
\begin{equation}
\frac{d\tau}{dz_d}(\sigma,\kappa) = \frac{dn}{d\sigma}\Delta \sigma  \frac{c\pi}{H_0} \frac{dH_0t}{dz}|_{z=z_d}(1+z_d)^3\left(max(\zeta,\xi_{mult})^2-\xi_{mult}^2\right).
\end{equation}

\subsection{the distribution of $\kappa$ for multiply-imaged sources}

\noindent The mapping $\zeta(\xi)$ has 1 or 3 solutions. As $\xi$ is decreased from large values to zero, $\zeta(\xi)$ crosses $\zeta=0$ , reaches $\zeta=-\xi_{mult}$ when ($\xi=\xi_{crit}$), and then increases to $\zeta=0$. This results in three cases.
The probability that a beam from a multiply-imaged source will pass through a microlensing optical depth smaller than $\kappa$ due to a galaxy halo with a velocity dispersion between $\sigma$ and $\sigma+\Delta\sigma$ and a redshift between $z_d$ and $z_d+\Delta z_d$ for each $\sigma$ is 
\begin{equation}
\frac{d\tau}{dz_d}(\sigma,\kappa) = 2\frac{dn}{d\sigma}\Delta \sigma   \frac{c\pi}{H_0} \frac{dH_0t}{dz}|_{z=z_d}(1+z_d)^3\left\{\begin{array}{ll}
\left(\xi_{mult}^2-min(\zeta,\xi_{mult})^2\right) & \mbox{for}\:\:\zeta>0\\
\left(\xi_{mult}^2+max(|\zeta|,\xi_{mult})^2\right) & \mbox{for}\:\:\zeta<0\:\:\mbox{and}\:\:\xi_{\kappa}>\xi_{crit}\\
\left(3\xi_{mult}^2-max(|\zeta|,\xi_{mult})^2\right) & \mbox{for}\:\:\zeta<0\:\:\mbox{and}\:\:\xi_{\kappa}<\xi_{crit}.\end{array}\right.
\end{equation}

\subsection{variation with redshift and magnification bias}

\begin{figure*}
\epsscale{1.}
\plotone{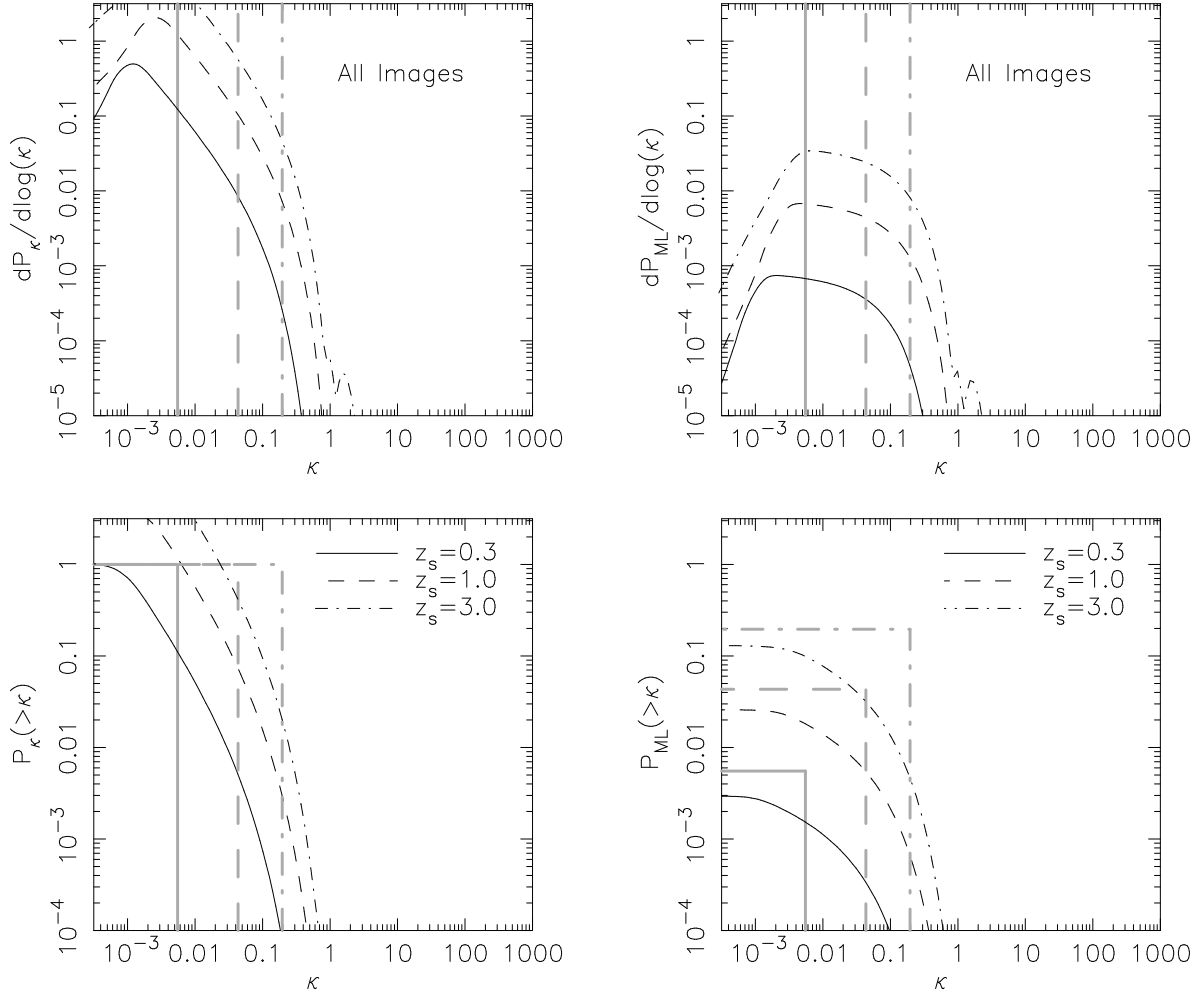}
\caption{The dependence of the probability of microlensing by dark compact objects on source redshift: Left: Differential (top) and cumulative (bottom) probabilities for the microlensing optical depth $\kappa$. Right:  Conditional differential (top) and cumulative (bottom) probabilities for the $\kappa$ along lines of sight near microlensed images. $\Omega_{co}$ was 0.3, and no magnification bias was assumed. Functions are shown for three values of $z_s$: 0.3 (solid lines), 1.0 (dashed lines) and 3.0 (dot-dashed lines). For comparison, the light lines show the corresponding probabilities where $\Omega_{co}$ is uniformly distributed.}
\label{NFW_Zs} 
\end{figure*}

\noindent $\frac{dP_{\kappa}}{d\kappa}$, $\frac{dP_{ML}}{d\kappa_{ML}}$, $P_{\kappa}(>\kappa)$ and $P_{ML}(>\kappa_{ML})$ for all sources were found from the pairs of differential and cumulative distributions calculated for singly and multiply-imaged sources. Fig.~\ref{NFW_Zs} shows these for sources at redshifts $z_s=$ 0.3, 1.0 and 3.0. No magnification bias was assumed. 

The NFW profile is flatter than the isothermal sphere in the central regions, which reduces its relative cross-section, but increases the magnification of multiple images. Magnification bias may therefore be more important. For a circularly symmetric lens the magnification is
\begin{equation}
\mu = \left(\frac{D_s}{D_d}\right)^2\frac{\xi}{\eta}\left(\frac{d\eta}{d\xi}\right)^{-1}.
\end{equation}
The magnification of an image can therefore be calculated for the NFW profile from Eqn.~\ref{NFWeqn}.
Fig.~\ref{NFW_BIAS} shows $\frac{dP_{\kappa}}{d\kappa}$, $\frac{dP_{ML}}{d\kappa_{ML}}$, $P_{\kappa}(>\kappa)$ and $P_{ML}(>\kappa_{ML})$ including magnification biases calculated assuming 3 different values of $\alpha_m$: 0.0 (unbiased, solid lines), 0.2 (dashed lines) and 0.5 (dot-dashed lines). Plots are shown for sources at redshift $z_s=3.0$.

\section{Results and Discussion}
\label{results}

\noindent Probability distributions have been computed for the microlensing optical depth $\kappa$ due both to microlensing by stars in galaxies, and by compact objects in the dark halos around galaxies. Multiply-imaged sources are generally subject to larger microlensing optical depth, though there is some overlap in the distributions for singly and multiply imaged sources for both the cases of microlensing by stars and by compact objects in CDM halos. 

First we discuss some results for the probability of microlensing by stars.
Probability distributions were constructed for both elliptical/S0 and spiral galaxy populations. For ellipticals we find that larger cosmological stellar densities increase the probability of each $\kappa$, sometimes by a large amount. For example, the probability of each $\kappa$ is increased by around an order of magnitude between $\Omega_*=0.001$ and $\Omega_*=0.020$, though the increase is not uniform over all $\kappa$ due to the factor of $1/|1-\kappa_c|$. Naturally, this results in an overall increase in the probability of microlensing. In addition, at higher densities microlensing is more likely to be observed for sources with higher $\kappa_{ML}$'s near the line of sight (the distribution mode increases with $\Omega_*$). Conversely, our model of spiral galaxies retained a constant mass-to-light ratio in the disc, but varied the mass-to-light ratio in the bulge. The differential probability for $\kappa_{ML}<0.1$ is nearly independent of $\Omega_*$, suggesting that nearly all microlensing at these optical depths will be due to microlenses located in the disc. However there is some variation with $\Omega_*$ for $\kappa_{ML}>0.1$. In particular, multiply imaged sources have distributions that depend on $\Omega_*$, suggesting that most multiple images created by spiral galaxies are located in the bulge. 

It is rare for a source to be microlensed by stars. Between about 1 percent (for $\Omega_*=0.001$) and 10 percent (for $\Omega_*=0.020$) of high redshift sources are subject to $\kappa>0.01$ in stars in elliptical galaxies. As a result, only a few 10ths of 1 to 1 percent of high redshift sources will be microlensed by stars in elliptical galaxies at any one time. About 1 percent of high redshift sources are subject to $\kappa>0.01$ in stars in spiral galaxies. In our model, this number is nearly independent of $\Omega_*$. The resulting microlensing rate for high redshift sources is  around 1 10th of 1 percent, comparable to that in elliptical galaxies for $\Omega_*=0.001$, but lower otherwise.

The fraction of sources that are multiply-imaged by elliptical galaxies is only around 1 percent, and the fraction multiply-imaged by spiral galaxies is a factor of $\sim10$ smaller. However the larger $\kappa$'s near the lines of sight to multiple images means that in most cases where microlensing by stars is observed, the source will be multiply-imaged. For spiral galaxies, about 70 percent of microlensing will be observed in multiple images, while for ellipticals the discrepancy can be much larger, particularly if $\Omega_*$ is small. Note that the inclusion of angular and flux ratio resolution biases will lower the multiple imaging rate. This will result in a microlensing rate that is higher in single than multiple images in spirals, and in ellipticals if $\Omega_*$ is large. 

The remainder of the discussion on microlensing by stars refers to combined statistics for ellipticals and spirals. The probability distribution for $\kappa$ due to stars is sensitive to the source redshift. Each value of $\kappa$ is 10-20 times more likely along a random line of sight to a source at red-shift 3.0 than to a source at redshift 0.3. As a result, sources at redshift 3.0 will be microlensed at a rate more than 100 times that of sources at redshift $0.3$. On the other hand, we find that the most likely value of $\kappa$ ($\sim1$) is insensitive to source redshift.

To consider the effect of different cosmologies on the rate of microlensing by stars, we assumed a flat universe, and varied the cosmological constant $\Lambda$. Changing the cosmology does not change the form of the probability distribution for $\kappa$, or the value of its mode but does change the overall microlensing rate. Increasing $\Lambda$ from 0.0 to 0.9 increases the likely-hood of finding a given $\kappa$ along a random line of sight by a factor of 5-10. In addition, the chance that a source at redshift 3.0 is microlensed rises from $\sim0.0015$ if $\Lambda=0$ to $\sim 0.025$ if $\Lambda=0.9$.

Turning our attention to microlensing in CDM halos, we find that in contrast to microlensing by stars, all sources at redshift 0.3 are on average subject to normalized surface mass densities of $\frac{\Sigma}{\Sigma_{crit}}> 0.0005$, and all sources at redshift 3.0 to $\frac{\Sigma}{\Sigma_{crit}}>0.02$. If the dark matter is composed of compact objects, then this results in a microlensing rate for high redshift sources of more than 10 percent. The NFW profile describes halos with cores that are flatter than those of isothermal profiles. The rate of multiple imaging is therefore lower, and this combined with the more extended distribution of mass (compared with the de Vaucouleurs profile) results in microlensing by compact objects in NFW halos being dominated by singly rather than multiply imaged sources. In projection, galaxy halos are likely to have some overlap, and a source may be viewed through more than one. However, if more than one halo is important, the likely values of $\kappa$ are low and so the approximations employed in the calculations of the microlensing probabilities should remain valid.

\begin{figure*}
\epsscale{1.}
\plotone{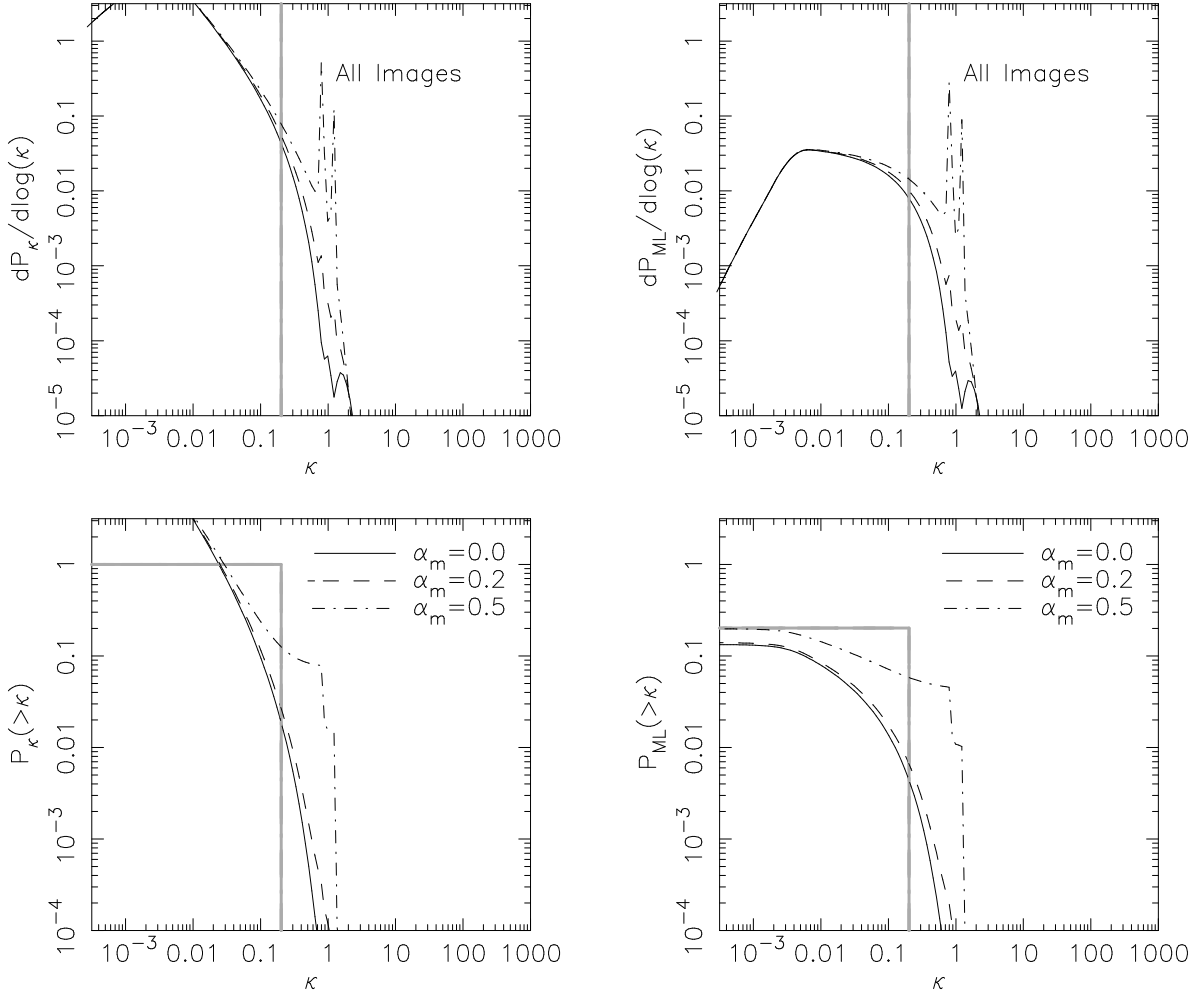}
\caption{The dependence of the probability of microlensing by dark compact objects on magnification bias: Left: Differential (top) and cumulative (bottom) probabilities for the microlensing optical depth $\kappa$. Right:  Conditional differential (top) and cumulative (bottom) probabilities for the $\kappa$ along lines of sight near microlensed images. $\Omega_{co}$ was 0.3, and a source redshift of $z_s=3.0$ was assumed. Functions are shown for three values of $\alpha_m$: 0.0 (unbiased, solid lines), 0.3 (dashed lines) and 0.5 (dot-dashed lines). For comparison, the light lines show the corresponding probabilities where $\Omega_{co}$ is uniformly distributed.}
\label{NFW_BIAS} 
\end{figure*}

The effect of magnification bias on the probability distribution for $\kappa$ was explored for both microlensing by stars and by objects in dark halos. We assumed a powerlaw form for the source luminosity function with an index $\alpha_m$ (Eqn.~\ref{numbercount}). Steeper forms of the luminosity function boost the number of multiple images relative to single images, and for microlensing by stars we find the curious result that the increased magnification bias reduces the number of images subject to large $\kappa$. This can be attributed to the fact that the highest values of $\kappa$ occur for image positions inside the Einstein radius, where the image magnification drops substantially. In the case of microlensing by stars, the most likely value of $\kappa$ near the line-of sight is insensitive to $\alpha_m$. However, an $\alpha_m$ of 0.5 reduces the total amount of microlensing by a factor of two, which reflects the importance of the contribution of multiple images to the statistics of microlensing by stars. 
For CDM halos, where $\kappa(\xi)$ is monotonic and the magnification of the central image is finite, large values of $\kappa$ are significantly more common as a result magnification bias. Because we have assumed an NFW profile, which has a low central density compared with an isothermal sphere, there are few multiply imaged sources. However these sources are significantly magnified. Luminosity functions with large values of $\alpha_m$ therefore significantly increase the likelihood of observing a source through a region of large microlensing optical depth. 
This can be seen from the spikes in the distributions in Fig.~\ref{NFW_BIAS} (for $\alpha_m=0.5$). In this case (source at redshift 3), the probability of a large value of $\kappa$ can be increased more than 100 times in the presence of a magnification bias due to a luminosity function with a large $\alpha_m$. The spikes in the distribution may be understood as follows. In the absence of magnification bias, radial depletion of images occurs near the Einstein radius where the source position lies near the optical axis, and at the critical radius for multiple imaging where radial arcs form. At both radii the large magnifications associated with the depletions result in local probabilities that are very sensitive to the source luminosity function. Furthermore, the Einstein radius and the critical radius for multiple imaging are found in the region where the profile is approximately isothermal. The values of $\kappa$ near these radii are therefore nearly constant with $\sigma$, resulting in the narrow peaks seen in Fig.~\ref{NFW_BIAS}. The increase in the total probability for microlensing due to magnification bias ($\alpha_m$) is only about 50 percent, demonstrating that multiple images are not a significant contributer to the microlensing statistics for NFW halos.

The microlensing optical depth can be computed for objects distributed randomly in co-moving volume (Press \& Gunn~1973; Turner, Ostriker \& Gott~1984). For comparison with our probability distributions, we have therefore computed the values of $\kappa_{unif}$ that result from the redistribution of stellar and galactic halo mass uniformly in space. For uniformly distributed matter the differential distributions for microlensing optical depth analogous to those computed in this paper are delta functions at $\kappa_{unif}$, and the resulting cumulative distributions are step functions. These distributions are plotted (light lines) in Figs.~\ref{SISell1}-\ref{NFW_BIAS}. In both the cases of microlensing by stars and by dark matter, the clustering results in a probability distribution for $\kappa$ whose mode is smaller than $\kappa_{unif}$. This is expected since in the limit of high clustering, all lines of sight have zero optical depth to microlensing. However, the distributions have significant probability spread over several orders of magnitude in $\kappa$, including a long tail extending to large $\kappa$. As a result, microlensing by stars in galaxies is much more likely ($>90$ percent) to be observed in images that have $\kappa$'s near the line of sight larger than $\kappa_{unif}$ (with a most likely value around 1). The coincidence of multiple imaging and large values of $\kappa$ (which doubles the chances of observing an image with $\kappa$), as well as the presence of smooth (dark) matter results in a microlensing rate for stars that is about a factor of 3 larger than that due to a uniform microlens distribution. In contrast, microlensing in galactic dark matter halos is most likely to be observed with $\kappa$'s near the line of sight that are smaller than $\kappa_{unif}$, although the mode lies atop a very broad peak. This results in a fraction of microlensed sources that (depending on magnification bias) is lower by nearly a factor of 2 than for uniformly distributed microlenses.

\section{Summary}

\noindent Probability distributions were computed for the microlensing optical depth $\kappa$ due both to microlensing by stars in galaxies, and, assuming dark matter to be in the form of compact objects, by microlenses in the dark halos around galaxies. These distributions were contrasted with the values of $\kappa$ in universes having uniformly distributed microlenses. Around 1 percent of high-redshift sources are microlensed by stars at any one time. In keeping with the strong lensing rate, incidences of microlensing by stars will be higher in elliptical/S0 than spiral galaxies. Interestingly, multiply-imaged sources dominate the stellar microlensing statistics, with less than 50 percent of microlensing by stars being of singly imaged (by the galaxy) sources. Our model suggests that in spiral galaxies, about 30 percent of microlensing will be in single images, and that these will be mostly located in the disk, with microlensed multiple images being located in the bulge.
 However if CDM halos are comprised of compact objects, then more than $\sim1$ high redshift source in 10 is microlensed at any one time. In addition, the vast majority of these sources are not multiply-imaged. Where compact objects are distributed in dark galactic halos, the simple calculation of microlensing probability from uniformly distributed objects provides reasonable results (to within a factor of 2). Furthermore, the typical $\kappa$ near lines of sight to microlensed images in dark matter halos will usually be comparable to or lower than the uniform value. In contrast, for microlensing by stars the typical optical depth is significantly larger than in the uniform case. Even more importantly, microlensing by stars will usually be observed at optical depths of order 1.

\acknowledgements
This research was supported by NSF grant AST98-02802 to ELT. JSBW acknowledges the support of an Australian Postgraduate Award.


\clearpage

\begin{thebibliography}{}

\bibitem[]{}
Bartelmann, M., 1996, A \& A, 313, 697

\bibitem[]{}
Bartelmann, M., 2000, A \& A, 357, 51

\bibitem[]{}
Boyle, B.J., Shanks, T., Peterson, B.A., 1988, MNRAS, 235, 935

\bibitem[]{}
Bryan, G., Norman, M. 1998, ApJ., 495, 80

\bibitem[]{}
Bullock, J.S., Kolatt, T.S., Sigad, Y., Somerville, R.S., Kravtsov, A.V., Klypin, A.A., Primack, J.R., Dekel, A., 2001, MNRAS, 351, 559

\bibitem[]{}
Carrol, S.M., Press, W.H., Turner, E.L., 1992, Annu. Rev. Astron. Astrophys., 30, 499

\bibitem[]{} 
Chang, K., Refsdal, S., 1979, Nature, 282, 561 

\bibitem[]{} 
Corrigan et al., 1991, Astron. J., 102, 34


\bibitem[]{}
Djorgovski, S., Davis, M., 1987, ApJ., 313, 59

\bibitem[]{}
Dalcanton, J.J., Canizares, C.R., Granados, A., Steidel, C.C., Stocke, J.T., 1994, ApJ., 424, 550

\bibitem[]{}
Faber, S., Jackson, R., 1976, ApJ., 204, 668

\bibitem[]{}
Fukugita, M., Futamase, T., Kasai, M., 1990, MNRAS, 246, 24 

\bibitem[]{}
Gott, J.R. III, 1981, ApJ., 243, 140

\bibitem[]{}
Gott, J.R. III, Gunn, J.E., 1974, ApJ., 190, 105

\bibitem[]{}
Hartwick, F.D.A., Schade, D., 1990, ARA\&A, 28, 437

\bibitem[]{}
Hawkins, M.R.S., 1993, Nature, 366, 242

\bibitem[]{} 
Irwin, M. J., Webster, R. L., Hewitt, P. C., Corrigan, R. T., 
Jedrzejewski, R. I., 1989, Astron. J., 98, 1989

\bibitem[]{}
Keeton, C.R., Kochanek, C.S., 1998, 495, 157

\bibitem[]{}
Kochanek, C.S., 1996, ApJ., 466, 638

\bibitem[]{}
Koopmans, L.V.E, Wambsganss, J., 2001, MNRAS, 325, 1317

\bibitem[]{}
Kuzmin, K., 1956, AZh, 33, 27

\bibitem[]{}
Li, L.-X., Ostriker, J.P, ApJ., Submitted, astro-ph/0010432 


\bibitem[]{}
Marani, G.F., Nemiroff, R.J., Norris, J.P., Hurley, K., Bonnell, J.T., 1999, ApJ., 512, L13

\bibitem[]{}
Marzke, R.O., Geller, M.J., Huchra, J.P. \& Corwin, H.G. Jr., 1994, Astron. J., 108, 437 


\bibitem[]{}
Metcalf, B.R., Silk, J., 1999, ApJ., 519, L1

\bibitem[]{} Minty, E.M., Heavens, A.F., Hawkins, M.R.S., 2001, MNRAS, submitted, astro-ph/0104221

\bibitem[]{}
M$\ddot{\mbox{o}}$rtsell, E., Goodbar, A., Bergstr$\ddot{\mbox{o}}$m, L., astro-ph/0103489

\bibitem[]{}
Navarro, J.F, Frenk, C.S., White, S.D.M., 1995, MNRAS, 275, 720

\bibitem[]{}
Navarro, J.F, Frenk, C.S., White, S.D.M., 1996, ApJ., 462, 563

\bibitem[]{}
Navarro, J.F, Frenk, C.S., White, S.D.M., 1997, ApJ., 490, 493


\bibitem[]{}
Netterfield, C.B., Ade, P.A.R., Bock,  J.J., Bond, J.R., et al., 2001, ApJ, submitted, astro-ph/0104460

\bibitem[]{}
Press, W.H., Gunn, J.E., 1973, ApJ., 185, 397


\bibitem[]{}
Tully, R.B., Fisher, J.R., 1977, A \& A, 54, 661

\bibitem[]{}
Turner, E.L., Ostriker, J.P., Gott, J.R. III, 1984, ApJ., 284, 1

\bibitem[]{}
Turner, E.L., 1980, ApJ., 242, L135

\bibitem[]{}
Turner, E.L., 1990, ApJ., 365, L43

\bibitem[]{}
Wang, Y., ApJ., 525, 651

\bibitem[]{}
Young, P., 1981, ApJ., 244, 756

\end{thebibliography}
\end{document}